\documentclass[prc,twocolumn,epsfig,nofootinbib,floatfix,showpacs]{revtex4-2}
\usepackage{graphics}
\usepackage{epsfig}
\usepackage{amsfonts}
\usepackage{amsmath}
\usepackage{bm}% bold math
\usepackage{color}

\begin{document}

\title{Microscopic analysis of M1 scissors mode in $^{254}$No}

\author{V.O. Nesterenko $^{1,2}$, M.A. Mardyban $^{1,2}$ and A. Repko $^{3}$}
\affiliation{$^1$
Laboratory of Theoretical Physics, Joint Institute for Nuclear Research.
141980, Dubna, Moscow region, Russia}
\affiliation{$^2$
Dubna State University. 141982, Dubna, Moscow region, Russia}
\affiliation{$^3$
Institute of Physics, Slovak Academy of Sciences, 84511 Bratislava, Slovakia}

\date{\today}

\begin{abstract}
The low-energy $M1$ orbital scissors mode (SM)  was recently observed by Oslo group
in deformed nucleus $^{254}$No. This is the heaviest nucleus where SM was ever
experimentally found. We propose the analysis of SM, together with the spin-flip
$M1$ resonance, within fully self-consistent Quasiparticle Random-Phase Approximation (QRPA) with
Skyrme forces SG2, SLy4 and SLy5. The impact of "tensor" $J^2$-term, introduced by
perturbative (on the base of SG2) and consistent (SLy5) ways, is analyzed and shown to be
noticeable but not decisive. The deformation-induced coupling of $M1$ and $E2$ states is
inspected. The calculations reasonably describe  Oslo's experimental data.
The best agreement is obtained for SLy5. A fine structure of SM in $^{254}$No is predicted.
A significant constructive interference of the dominant orbital and minor spin-flip contributions
to $M1$ strength at SM energy region is found. What is remarkable, our analysis of distributions of the
convective nuclear currents challenges the scissors-like flow usually assumed for SM.
%Instead, the mode looks rather as a lateral rotation-like oscillations concentrated  at
%Jthe equatorial regions of the nucleus.
\end{abstract}
\maketitle

%%%%%%%%%%%%%%%%%%%%%%%%%%%%%%%%%%%%%%%%%%%%%
%%%%%%%%%%%%%%%%%%%%%%%%%%%%%%%%%%%%%%%%%%%%%
\section{Introduction}

Nobelium isotopes lie  at the edge of superheavy elements and so their description
can serve as a paramount test for theoretical models pretending for exploration of superheavy nuclei.
Furthermore, these isotopes are the most heavy nuclei where we still have  substantial
experimental spectroscopic data \cite{Herz08,Hess23,Acker25}. There is a
bulk of theoretical work devoted to nobelium isotopes, see e.g. recent studies
\cite{Dob15,Guo24,Xu24,Dao24,Balb24,Nest25} and references therein. In our previous work
\cite{Nest25}, we reported a systematic analysis of ground-state properties and multipole electric
excitations in $^{250-262}$No in the framework of self-consistent Quasiparticle Random-Phase
Approximation (QRPA) method with Skyrme forces. The main attention was paid to $^{252,254}$No, where
the structure of K-isomers and recent experimental data \cite{Forge23,Wahid25} for low-energy monopole
and hexadecapole states were successfully described. A significant shell gap in the neutron
single-particle spectrum and corresponding reduction of the neutron pairing in  $^{252,254}$No
were predicted.

In the recent Oslo experiment $^{208}$Pb($^{48}$Ca, 2n$\gamma$)$^{254}$No, the first evidence of $M1$
scissor mode (SM) in $^{254}$No was reported~\cite{Garotte}. In this connection, we propose
Skyrme QRPA analysis of $M1$ states in $^{254}$No, including SM and nearby spin-flip excitations.

The $M1$ scissors mode was first predicted within two-rotor model~\cite{Iud78} and then observed
in high-resolution $(e,e')$ experiment~\cite{Boh84}. This collective
mode is usually treated as  out-of-phase orbital oscillations of proton and neutron deformed subsystems.
On the macroscopic level, this mode was also suggested in sum-rule approach~\cite{Lip83,LSPR89}, where the coupling
between SM and high-energy isovector branch $E2(K=1)$ of the isovector quadrupole giant resonance (IV-GQR)
was taken into account.  SM is a remarkable manifestation of nuclear orbital magnetism.
The mode is represented by $K^{\pi}=1^+$ states ($K$ is projection of the total nuclear momentum into
the symmetry z-axis and $\pi$ is the parity) lying at 2.5--3.5 MeV and demonstrating a
significant orbital M1 strength. The SM is also interesting as an example of mixed-symmetry
states in well-deformed nuclei~\cite{Hei10}.

The features of SM were a subject of intensive experimental and theoretical studies,
see reviews~\cite{Rich95,Iud97,Hei10}. By construction, SM should exist in all deformed nuclei
and, indeed, it was observed in deformed light/medium ($A < 140$), rare-earth
and actinide nuclei, see experimental systematics~\cite{Enders05}. However, the search of
SM in very heavy nuclei like $^{254}$No is a challenging task due to short lifetimes
and high level densities. In Oslo experiment
$^{208}$Pb($^{48}$Ca, 2n$\gamma$)$^{254}$No, the  $\gamma$-spectroscopy techniques was
applied to identify SM as a hump at $\sim$ 2.5 MeV with the summed
$\sum B_{\rm SM}(M1\uparrow)$=11.8(19) $\mu_N^2$~\cite{Garotte}.

The paper~\cite{Garotte} also offers  theoretical estimations
for SM in $^{254}$No, using the sum-rule approach~\cite{Enders05}, QRPA method
with Gogny forces ~\cite{Gogny} and some phenomenological
expressions~\cite{Mump17,Gor14}. The sum-rule estimations ($E_{\rm SM}$=2.1 MeV,
$\sum B_{\rm SM}(M1\uparrow)$=12.1(13) $\mu_N^2$) are rather close to experimental values.
The QRPA-Gogny results, despite a significant ad hoc energy downshift, demonstrate
a worse performance. The phenomenological evaluations strongly overestimate~\cite{Mump17}
or underestimate~\cite{Gor14} the strength $\sum B_{\rm SM}(M1\uparrow)$. In addition,
the recent study~\cite{Balb24} within Wigner-Function-Moments
(WFM) method gives the  reasonable results $E_{\rm SM}$= 2.60 MeV and
$\sum B_{\rm SM}(M1\uparrow)$=13.4 $\mu_N^2$. However, WFM
 uses a simplified mean-field potential (anisotropic harmonic oscillator)
and is not self-consistent. Altogether, despite some first estimations, there is an
obvious need in a further  exploration of SM in $^{254}$No, in particular,
 within modern self-consistent microscopic models.

In this work, we present a microscopic analysis of $M1$ response in $^{254}$No
using fully self-consistent QRPA model~\cite{Repko_QRPA} with Skyrme forces SG2~\cite{Sg2},
SLy4~\cite{SLy5} and SLy5~\cite{SLy5}. We take into account the impact of so-called
$J^2$-term which can include central-exchange and tensor contributions~\cite{Colo07}.
Following our~\cite{Ves_PRC09,Nest_JPG10,Nest_EPJA24} and other~\cite{Zal_tensor,Bender_tensor,Cao_tensor}
studies, this term can be important for description of M1 excitations. For simplicity,
we limit ourselves by the central-exchange contribution to $J^2$-term. The Skyrme force
SG2~\cite{Sg2} is chosen since it was earlier successfully used for description
of various $M1$ modes, including SM and spin-flip giant resonance
(SFGR)~\cite{Ves_PRC09,Nest_JPG10,Nest_EPJA24,Pai16}.
Here SG2~\cite{Sg2} is applied  without and with $J^2$-term
(versions SG2 and  SG2+$J^2$, respectively). Besides we employ the
forces SLy4~\cite{SLy5} (without $J^2$-term)
and SLy5~\cite{SLy5} ($J^2$-term with the refit of the parameters).

Unlike the previous studies of $M1$ strength in $^{254}$No, we analyse not only energy spectra and $M1$ strengths
but also distributions of the convective nuclear current in SM states. This allows to visualize
the actual flow of protons and neutrons, as well as their isoscalar (IS) and isovector (IV)
patterns, in SM energy region. Besides, we investigate the interference of
the orbital and spin contributions  to $M1$ states and deformation-induced coupling
of $M1(K=1)$ and $E2(K=1)$ excitations.
% ($K$ is projection of the spin on the symmetry z-axis).

 The paper is organized as follows. In Section II, the model and details of the calculations are sketched.
 In Section III, the results and discussion are presented. The conclusions are drawn in Section IV. In Appendix A,
 the spin-orbital and tensor parameters are defined.

\section{Model and calculation details}

The calculations are performed within fully self-consistent matrix QRPA
model~\cite{Repko_QRPA,Repko_EPJA17_pairing,Kvasil_EPJA19_SEBRPA,Repko_PRC19_SEARPA}
based on Skyrme functional~\cite{Bender_RMP03}. The self-consistency means that:
 (i) both mean field and residual interaction
are derived from the same Skyrme functional, (ii) the contributions
of all time-even densities and time-odd currents from
the functional are taken into account, (iii) both particle-hole
and pairing-induced particle-particle channels are included,
(iv) the Coulomb (direct and exchange) parts are involved
in both mean field and residual interaction. Spurious admixtures
caused by violation of the rotational invariance are removed using the
technique~\cite{Kvasil_EPJA19_SEBRPA}. The pairing is treated within the Bardeen-Cooper-Schrieffer
(BCS) scheme using the zero-range surface pairing interaction~\cite{Repko_EPJA17_pairing}.

As mentioned above,  the representative set of Skyrme forces (SG2~\cite{Sg2}, SLy4~\cite{SLy5}
and SLy5~\cite{SLy5}) is used.
In Table~\ref{tab-1}, some characteristics of the forces are shown.
In particular, we exhibit: IS effective masses $m_0/m$ which can significantly affect
the single-particle spectra~\cite{Nest_PRC04}; spin-orbit parameters $b_4$ which, in a large extent,
determine the energy and structure of SFGR~\cite{Ves_PRC09,Nest_JPG10};
tensor parameters $\tilde{b}_1$ and $\tilde{b}'_1$  which define $J^2$-term.
%proton and neutron pairing constants $G_p$ and neutron $G_n$.
The expressions for spin-orbit and
tensor parameters through  the standard Skyrme parameters are given in Appendix A.
Table~\ref{tab-1} shows a big difference between SG2 and SLy4-SLy5 parameters.
%Despite the refit of SLy5 parameters after inclusion of $J^2$ term,
The values $b_4$, $\tilde{b}_1$, and $\tilde{b}'_1$ for SLy4 and SLy5
are rather similar despite the refit in SLy5.

\begin{table} %Table 1
\centering
\caption{IS effective masses  $m_0/m$,  spin-orbit parameters $b_4$, and tensor
parameters $\tilde{b}_1$ and $\tilde{b}'_1$ for SG2, SLy4 and SLy5.}
\label{tab-1}
\begin{tabular}{|c|c|c|c|c|}
\hline
 force & $m_0/m$ & $b_4$ & $\tilde{b}_1$ & $\tilde{b}'_1$  \\
       &         & MeV $\rm{fm}^5$ & MeV $\rm{fm}^5$ & MeV $\rm{fm}^5$  \\
 \hline
 SG2  & 0.79 & 52.5 & -9.96 & -47.74 \\
 SLy4 & 0.70 & 61.5 &  47.37    &  -129.15      \\
 SLy5 & 0.70 &  63.0 & 48.87   & -129.07 \\
\hline
\end{tabular}
\\
\end{table}

\begin{table} % Table 2
\centering
\caption{Deformation parameters $\beta_2$ and $\beta_4$, proton $\Delta_p$ and
neutron $\Delta_n$ pairing gaps, energies $E_{2^{+}_1}$ of $2^+_1$ state of the
ground-state rotational band and binding energy per nucleon $BE/A$,
calculated with SG2, SG2+$J^2$, SLy4 and SLy5 for $^{254}$No. The
experimental data for $\beta_2$~\cite{Rei99,Herz01}, $E_{2^{+}_1}$~\cite{nndc} and
$BE/A$~\cite{nndc} are shown.}
\label{tab-2}
\begin{tabular}{|c|c|c|c|c|c|c|c|c|}
\hline
       & $\beta_2$  & $\beta_4$ & $\Delta_p$ & $\Delta_n$ &  $E_{2^{+}_1}$ & $BE/A$\\
       &         &            &  [MeV]       &   [MeV]      & [keV] & [MeV] \\
\hline
 exper & 0.27(3)-0.32(2) &           &          &           & 44 & 7.42 \\
\hline
 SG2   &  0.299  &    0.059     & 0.53 & 0.35 & 43 & 7.44 \\
\hline
 SG2 + $J^2$ &  0.296  &  0.062  & 0.54 & 0.59 & 49 & 7.42 \\
 \hline
SLy4         & 0.304  &  0.061  &   0.47 & 0.28 &  38 & 7.33 \\
\hline
SLy5         & 0.303  &  0.061  &   0.49 & 0.43 &  44 & 7.33\\
\hline
\end{tabular}
\\
\end{table}

The mean field spectra and pairing characteristics in axially deformed
$^{254}$No are calculated by the code SkyAx~\cite{skyax} using a two-dimensional
grid in cylindrical coordinates. The calculation box extends
up to three nuclear radii, the grid step is 0.7 fm. All
proton and neutron single-particle (s-p) levels from the bottom of the potential well
up to +40 MeV are taken into account. For example, SLy5 calculations
employ about 1800 proton and 2000 neutron s-p levels.

The axial quadrupole $\beta_2$ and hexadecapole $\beta_4$ equilibrium deformations are
obtained by minimization of the total energy of the system.
As seen from Table~\ref{tab-2}, the calculated $\beta_2$ for SG2, SG2+$J^2$, SLy4 and SLy5
lie between the experimental values 0.27(3)~\cite{Rei99} and 0.32(2)~\cite{Herz01}.
Besides, we see a significant hexadecapole deformation $\beta_4$.
The $J^2$-contribution (SG2+$J^2$) noticeably increases
the neutron pairing gap as compared to SG2 case.
%For exception of SG2+$J^2$ and SLy4,
The calculations well describe (with some deviations for SG2+$J^2$ and SLy4) the energy
$E_{2^{+}_1}$ of $2^+_1$ state of the ground-state rotational band.
This energy is determined by the moment of inertia, which in turn depends on the nuclear
deformation and pairing. A general good agreement with experiment for  $E_{2^{+}_1}$ means that
the applied Skyrme parametrizations generally well describe a balance between deformation and pairing
in $^{254}$No. The binding energies per nucleon $BE/A$ are somewhat underestimated for SLy4 and SLy5.
 In general, the best performance in Table~\ref{tab-2} is obtained for SG2.

\begin{figure*} % Fig. 3
\label{fig1:sf}
\centering
\includegraphics[width=22cm,angle=0, scale=0.8]{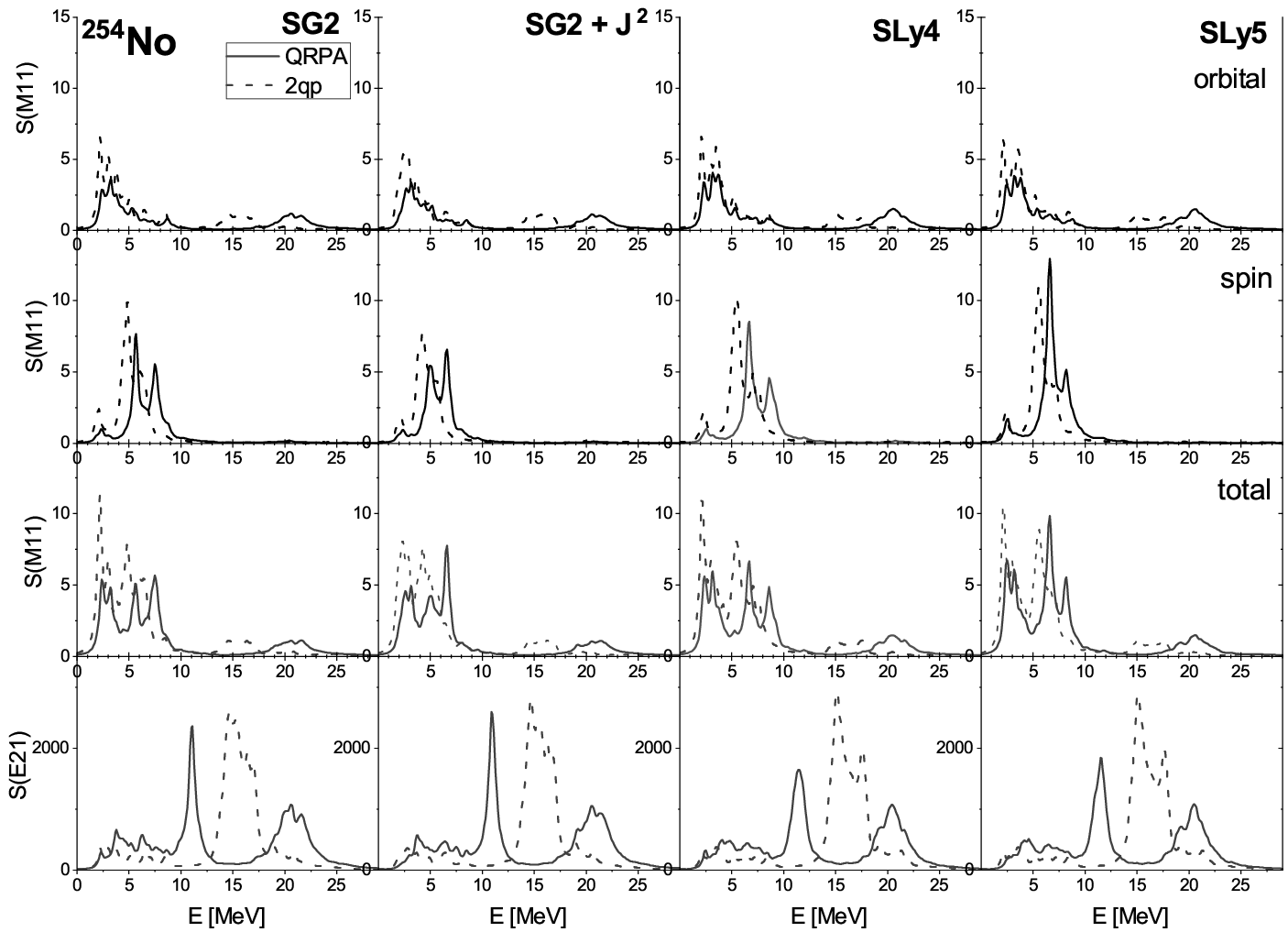}
\caption{QRPA (solid curves) and 2qp (dash curves) strength functions  in $^{254}$No,
calculated  with the forces SG2, SG2+$J^2$, SLy4 and SLy5.
$M11$ strength functions $S(M11)$ (in units $\rm{\mu^2_N/MeV}$) are shown for
orbital (first line), spin (second line) and total  (third line) cases.
In the bottom line, the quadrupole strength function $S(E21)$ (in units $\rm{e^2fm^4/MeV}$)
is exhibited. The strengths are smoothed by the Lorentz weight with the averaging
parameter $\Delta$=0.5 MeV.}
\end{figure*}

 The reduced probability for $M1(K=1)$ transitions
 ($M11$ in the short notation)
from the ground state $|0\rangle$ to the excited QRPA state $|\nu\rangle$ with
$I^{\pi}K=1^+1$ reads
\begin{equation}
\label{eq:BM11}
B(M11,\nu)=2|\:\langle\nu|\:\hat{\Gamma}(M11)\:|0\rangle \:|^2 .
\end{equation}
The coefficient  2 is used to take into account contributions of both
projections K=1 and -1. The transition operator is
\begin{equation}
\label{eq:M1}
 {\hat \Gamma}(M11) =
 \mu_N \sqrt{\frac{3}{4\pi}}\sum_{q \epsilon p,n}
 %\sum_{k \epsilon q}
[g^{q}_s {\hat s}_{\mu=1} + g^{q}_l {\hat l}_{\mu=1}] ,
\end{equation}
where $\mu_N$ is the nuclear magneton, $\hat{s}_{\mu=1}$ and $\hat{l}_{\mu=1}$ are
$\mu$=1 projections of the standard spin and orbital operators,
$g^{q}_s$ and $g^{q}_l$ are spin and orbital gyromagnetic factors. The spin
g-factors are $g^{q}_s = \eta \bar{g}^{q}_s$
where  $\bar{g}^{p}_s$ = 5.58  and  $\bar{g}^{n}_s$ =-3.82
are bare proton and neutron g-factors and $\eta$=0.7 is the familiar
quenching parameter \cite{Har01}. The orbital g-factors are  $g^{p}_l$ = 1
and  $g^{n}_l$ = 0. In this study, we consider three
relevant cases: spin ($g^{q}_l=0$), orbital ($g^{q}_s=0$), and total (where both
spin and orbital contributions are taken into account). The corresponding transition
probabilities are marked as $B_{\rm orb}(M11)$, $B_{\rm spin}(M11)$ and $B_{\rm tot}(M11)$,
respectively.
%The expressions for
%the orbital and spin $M11$ matrix elements are given in the Appendix B.

In deformed nuclei, electric and magnetic states with the same $K^{\pi}$
are mixed \cite{Har01}. For $K^{\pi}=1^+$ states, the  modes
of multipolarities $M11$ and $E21$ are mixed. To inspect this mixing, we also
calculate the reduced probability of $E21$ transitions
\begin{equation}
\label{BE21}
B(E21,\nu)=2|\:\langle\nu|\:\hat{\Gamma}(E21)\:|0\rangle \:|^2
\end{equation}
with the transition operator
\begin{equation}
\label{eq:E21}
 {\hat \Gamma}(E21) =
  e \sum_{q \epsilon p,n} e_{\text{eff}}^q
%\sum_{k \epsilon q}
r^2 Y_{21}(\theta,\phi) ,
\end{equation}
where $Y_{21}(\theta,\phi)$ is the spherical harmonic and $e_{\text{eff}}^q$ are
effective charges $e_{\text{eff}}^p$=1 and $e_{\text{eff}}^n$=0.
% in the standard case and
%$e_{\text{eff}}^p$=-$e_{\text{eff}}^n$=1 in IV case).

Using $B(M11,\nu)$ and $B(E21,\nu)$, we calculate strength functions
\begin{eqnarray}
 \label{SFM11}
 S(M11;E) &=& \sum_{\nu} B(M11,\nu)\zeta(E_{\nu}, E),
 \\
  S(E21;E) &=& \sum_{\nu}  B(E21,\nu)\zeta(E_{\nu}, E),
 \label{SFE21}
\end{eqnarray}
 where
$ \zeta(E_{\nu},E) = \Delta /[2\pi[(E- E_{\nu})^2+\frac{\Delta^2}{4}]]$
is a Lorentz weight with an averaging parameter $\Delta$.

The distributions of the nuclear current are described by the current
transition densities (CTD)
\begin{equation}
\delta \bold {j}_{\nu}(\bold{r}) = \langle \nu| \hat{\bold j}|0\rangle (\bold{r})
\label{CTD}
\end{equation}
for the convective nuclear current
\begin{equation}
\label{j_con}
\hat{\bold j} (\bold r)= -i \frac{e\hbar}{2m} \sum_{q =n,p}e_{\text{jeff}}^q
\sum_{k \epsilon q}(\delta({\bold r} - {\bold r}_k) {\overrightarrow{\bold \nabla}}_k
+ {\overleftarrow{\bold \nabla}}_k \delta({\bold r} - {\bold r}_k)) .
\end{equation}
Here $e_{\text{jeff}}^q$ are the effective charges. They are
$e_{\text{jeff}}^p$=1 and $e_{\text{jeff}}^n$=0 for the proton current,
$e_{\text{jeff}}^p$=0 and $e_{\text{jeff}}^n$=1 for the neutron current,
$e_{\text{jeff}}^p=e_{\text{jeff}}^n$=1 for IS current and
$e_{\text{jeff}}^p=-e_{\text{jeff}}^n$=1 for IV current.

\section{Results and discussion}
\subsection{Strength functions}

Figure 1 gives a general view of the strength distributions $S(M11)$ and $S(E21)$
in $^{254}$No, calculated with SG2, SG2+$J^2$, SLy4 and SLy5. To demonstrate both IS and IV
strengths, the strength function  $S(E21)$ is calculated with effective charges
$e_{\text{eff}}^p$=1 and $e_{\text{eff}}^n$=0. To illustrate the impact
of the residual interaction, the unperturbed 2qp and QRPA strength functions
are compared. For $M11$ response, the orbital, spin and total cases are depicted. The strengths
are smoothed by Lorentz weight with the averaging parameter $\Delta$=0.5 MeV.
%For a view convenience, the averaging parameter  $\Delta$=0.5 MeV in the Lorentz weight is used.

It is seen that QRPA $M11$ strength is somewhat upshifted relative to 2qp strength.
So, for $M11$ modes, the residual interaction is basically IV and is not strong. Instead,
in $E21$ case, the effect of both IS and IV residual interactions is essential. As a result,
$E21$ 2qp strength at $\sim$15 MeV  is  transformed into IS ($\sim$10 MeV) and IV ($\sim$20 MeV)
parts of the giant quadrupole resonance (GQR).

 The upper panels of Fig. 1 show that the orbital $M11$ strength is concentrated at 1-5 MeV,
 where it forms  the SM. Besides, there is a high-energy bump at 20 MeV. This bump correlates with IV-GQR
 (see the bottom panels), which manifests the deformation-induced coupling of $M11$ and $E21$ modes.
 Note that the high-energy part of the orbital strength plays the important role in the sum-rule estimations
 for SM~\cite{Enders05,Lip83,LSPR89}.

 The next panels show that spin-flip  $M11$ strength is dominated by two-hump IV SFGR located at 5-10 MeV.
 As demonstrated in Fig. 2, these humps are produced by proton (low-energy hump)
 and neutron (high-energy hump)  spin-flip transitions. The energies of these transitions are mainly
 determined by spin-orbital splitting in the proton and neutron single-particle schemes. In nuclei with
 $N \gg Z$, the neutron Fermi level usually lies in the region with higher orbital moments
 (as compared to the proton case), and so neutrons demonstrate a stronger spin-orbital
 splitting than protons. For this reason, the neutron spin-flip hump
 lies above the proton one. In Skyrme parametrizations, the magnitude of spin-orbit
 splitting is determined by the parameter $b_4$. As seen in Table~\ref{tab-1}, the forces SLy4
 and  SLy5 have a higher $b_4$ than SG2. As a result, SFGR for SLy4 and SLy5 lies at a higher energy than
 in SG2 and SG2+$J^2$ cases.

 The panels at the third line of Fig. 1 show that the total low-energy $M11$ strength is
 generally formed by adjoined SM
 (1-5 MeV) and SFGR (5-10 MeV). So one should expect a mixture of these two $M11$ modes at the
 energy around 5 MeV. Note that SG2, SG2+$J^2$, SLy4 and SLy5 demonstrate rather similar
strength distributions for SM but different for SFGR. The impact of $J^2$-term  is  essential
mainly for SFGR. As seen from the bottom panels, the SM  energy region includes also a
significant $E21$ strength.

\begin{figure} %Fig2
\label{SFGR}
\centering
\includegraphics[width=10cm,angle=0, scale=0.75]{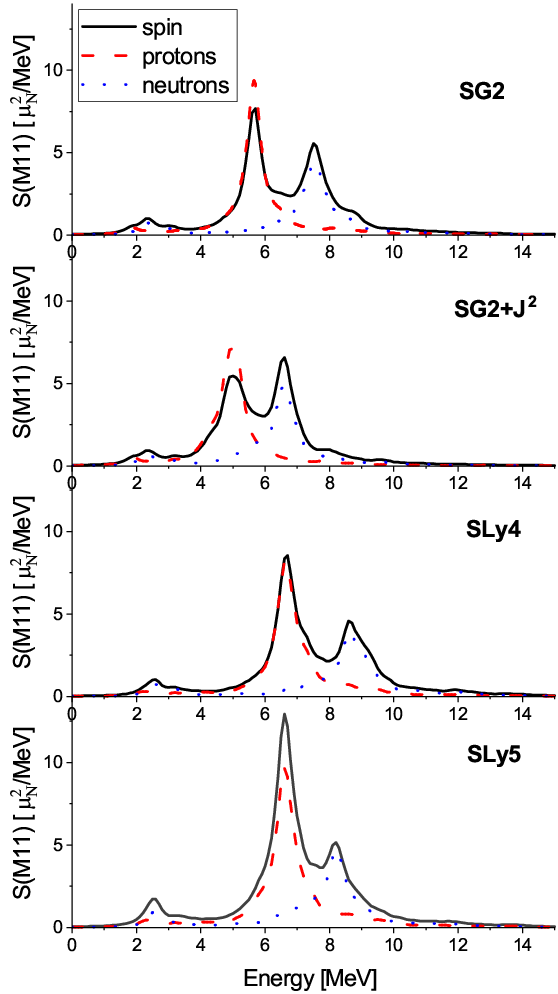}
\caption{Proton (red dashed line) and neutron (blue dotted line) branches  of $M11$ SFGR
as well as the total SFGR strength (black solid line) in $^{254}$No, calculated with SG2,
 SG2+$J^2$, SLy4 and SLy5.}
 %The units $\mu^2_N/$MeV are used.}
\end{figure}
\begin{figure*} %Fig. 3
\label{B(M11)+B(E21)}
\centering
\includegraphics[width=22cm,angle=0, scale=0.8]{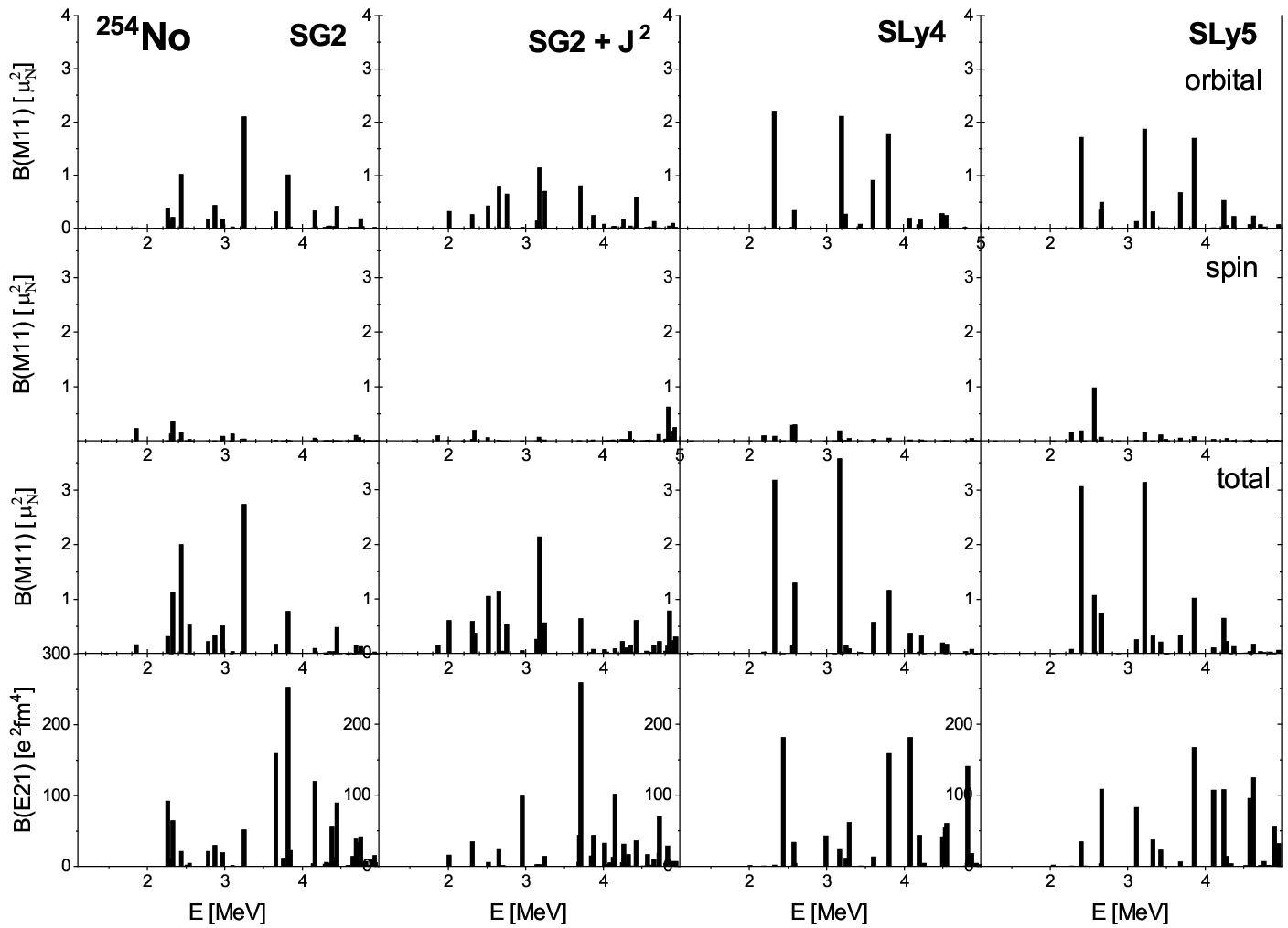}
\caption{Orbital (first line), spin (second line) and total (third line) $B(M11)$-values
in $^{254}$No for QRPA $K^{\pi}=1^+$ states at 1-5 MeV,  calculated with the forces
SG2, SG2+$J^2$, SLy4 and SLy5. In the bottom line, IV $B(E21)$ values are exhibited.}
\end{figure*}

\begin{table*} %Table 3
\caption{Characteristics of low-energy $1^+$ states with large values of $B(M11)$ in
$^{254}$No, calculated with SG2, SG2+$J^2$, SLy4 and SLy5. The table includes:
excitation energies $E$; orbital  $B_{\rm orb}(M11)$, spin $B_{\rm spin}(M11)$ and total $B_{\rm tot}(M11)$
reduced transition probabilities for $M11$ transition, IV quadrupole transition probability $B(E21)$,
main 2qp components $qq'$ (in terms of asymptotic Nilsson quantum numbers) with their energies
$\epsilon_{qq'}$, contributions $N_{qq'}$ to the state norm and locations relative to the
Fermi-level $F$.}
\begin{tabular}{|c|c|c|c|c|c|c|c|c|c|}
\hline
Force & E    &  \multicolumn{3}{c|}{$B(M11)$ [$\mu ^2_N$]}
  &  $B(IV,E21)$    & $qq'$ & $\epsilon_{qq'}$  & $N_{qq'}$  & F-location\\
 \hline
      & [MeV] & orbital & spin & total &  [$W.u.$] &    &[MeV]  & &\\
\hline
SG2 & 2.44 & 1.0 & 0.2 & 2.0 & 1.6 & pp[633$\uparrow$, 624$\uparrow$] & 2.22 & 0.60 & F-1, F+2\\
 &  & && && nn[622$\uparrow$, 613$\uparrow$] & 2.67 & 0.15 & F-1,F+4\\
 \hline
   & 3.25 & 2.1 & 0.04 & 2.7 &  1.6 &  pp[521$\uparrow$, 512$\uparrow$] & 3.06 & 0.78 & F-2, F+3\\
 &  & && && nn[624$\downarrow$, 615$\downarrow$] & 3.00 & 0.11 & F-2, F+3  \\
% \hline
% & 3.82 & 1.010 & pp[514$\downarrow$, 523$\downarrow$] & 3.81 & 0.48 \\
% &  &  & pp[512$\uparrow$, 521$\uparrow$]  & 3.84 &  0.22\\
 \hline
SG2+J$^2$ & 2.51& 0.4 & 0.1& 1.0 & 0.6 &  pp[633$\uparrow$, 624$\uparrow$] & 2.37 & 0.68 & F+1, F+3\\
&  & && && pp[505$\downarrow$, 514$\downarrow$] & 2.71 & 0.13 & F-1, F+4\\
%\hline
%& 2.75 & 0.652& pp[505$\downarrow$, 514$\downarrow$] & 2.71 & 0.58  \\
%&  &  & nn[613$\uparrow$, 622$\uparrow$] & 2.73 & 0.17 \\
\hline
& 3.18 & 1.1 & 0.2 & 2.1 &  1.3 &  pp[521$\uparrow$, 512$\uparrow$] & 2.94 & 0.45 & F-2, F+2  \\
&  & && && nn[631$\uparrow$, 620$\downarrow$] & 3.09 & 0.32 & F-2, F+2\\
\hline
SLy4 & 2.32 & 2.2 &  0.1 & 3.2 & 1.1 &  pp[633$\uparrow$, 624$\uparrow$]& 2.13 & 0.68 & F-1, F+2\\
&  & & &  & & nn[734$\uparrow$, 725$\uparrow$]  & 2.23 & 0.27 & F, F+4\\
\hline
& 3.17 & 2.2 & 0.2 & 3.6 &   1.6  &   pp[521$\uparrow$, 512$\uparrow$] & 3.01 & 0.71 & F-2, F+3 \\
&  & && && nn[622$\uparrow$, 613$\uparrow$] & 2.89 & 0.14 & F-2, F+3\\
\hline
SLy5 & 2.40 & 1.7 & 0.2 & 3.1 & 0.4 &  pp[633$\uparrow$, 624$\uparrow$] & 2.13 & 0.77 & F-1, F+2\\
&  & && && nn[743$\uparrow$, 734$\uparrow$] & 2.43 & 0.05 & F-3, F \\
\hline
& 3.22 & 1.9 & 0.2 & 3.1 & 0.003  &   pp[521$\uparrow$, 512$\uparrow$] & 3.01 & 0.71 & F-2, F+3\\
&  & && && nn[631$\downarrow$, 620$\uparrow$] & 3.32 & 0.16 & F-5, F+1\\
%&  3.83 & 1.574 & pp[512$\uparrow$, 521$\uparrow$] &  3.79 & 0.56 \\
%&  &  & nn[615$\downarrow$, 624$\downarrow$] & 3.41 & 0.15 \\
\hline
\end{tabular}
\label{tab-3}
\end{table*}

\subsection{SM energy region}

In Figure 3, the $M11$ (orbital, spin, total)  and $E21$ responses
are shown for separate QRPA $K^{\pi}=1^+$ states  at the energy  1-5 MeV,
where SM is located.
It is seen that in this energy region the orbital part
strongly dominates over the spin one. However, the spin part is important because
of the significant constructive interference between orbital and spin contributions
to the total $B_{\rm tot}(M11)$. This interference is clearly seen in Figure 3  and
Table~\ref{tab-3} for the states with large $B_{\rm tot}(M11)$ located at 2.0-3.5 MeV, where
$B_{\rm orb}(M11)+B_{\rm spin}(M11) < B_{\rm tot}(M11)$.

Note that each $1^+$ state at 1 $<$E$<$ 5 MeV exhibits both $B(M11)$ and $B(E21)$
responses, which is the result of deformation-induced $M11/E21$ coupling.
As seen from Fig. 3, the peaks with strong $B(M11)$ and $B(E21)$
strengths usually do not correlate. At the same time, a noticeable correlation
can be found between $B(M11)$ and IV $B(E21)$ ($e_{\text{eff}}^p$=-$e_{\text{eff}}^n$=1)
(not demonstrated here). Following Fig. 1, the quadrupole residual interaction is strong.
So, perhaps, the quadrupole fraction mainly affects the collectivity of $K^{\pi}=1^+$ states.
%Unlike the WFM conclusions~\cite{Balb_NPA03}, we have found no evidence that high-energy
%IV-GQR resonance affects the SM features.

Figure 3 shows  that distributions of the orbital and total $M11$ strengths
are different for SG2 and SG2+$J^2$  and rather similar for SLy4 and SLy5.
This is because $J^2$-term in SG2+$J^2$ is added by a
perturbative way, i.e.  without the corresponding refit of other parameters. Thus we get a
significant impact. Instead, introduction of $J^2$-term with the refit in SLy5
results in much weaker effect. So the refit of the parameters is important for the
proper estimation of $J^2$ impact.

Further, figure 3 demonstrates some fine structure in the distributions of $M11$ strengths.
Namely,  orbital and total $M11$ strengths are roughly separated into two-three
groups located at $\sim$2.5 MeV, $\sim$3.2 MeV and $\sim$3.8 MeV. The groups are
characterized by concentration of the strength or (and) by some high peaks with the
strength 1.5-3.5 $\mu ^2_N$. Our analysis shows that high peaks at $\sim$2.5  MeV and
$\sim$3.2 MeV can be explained by proton particle-hole transitions
$[633\uparrow] \to [624\uparrow]$  and  $[521\uparrow] \to [512\uparrow]$.
As seen from Table~\ref{tab-3},  just these transitions
%(with exception of 2.52-MeV state in SLy5 case)
dominate in QRPA states with a large total $B(M11)$.
The large $M11$ strength is achieved because these proton transitions
%These
%The states demonstrate large $B_{\rm tot}(M11)\sim 2-3 \mu ^2_N$ because the dominant
% proton 1ph transitions
  entirely fulfill selection rules for Nilsson asymptotic quantum numbers:
$[N n_z \Lambda]$:
$\Delta N =0,\pm 2;  \Delta n_z = 0,\pm 1; \Delta \Lambda = 0,1$\cite{Sol76}. Somewhat small
values of $B_{\rm orb}(M11)$ and $B_{\rm tot}(M11)$ for 2.51-MeV state in SG2+$J^2$ case
are explained by particle-particle $(F+1,F+3)$ character of this particular transition.
So, our calculations explain the SM fine structure mainly by particular prominent
1ph $M11$ transitions.

\begin{table*} % Table 4
\caption{Comparison of the experimental data~\cite{Garotte},
estimations~\cite{Garotte,Mump17,Balb24} and present
results for the centroid energies $E_c$ and summed $B(M11)$ (orbital, spin, total)
strengths, obtained at the energy intervals $\Delta$E.}
\begin{tabular}{|c|c|c|c|c|c|}
\hline
Method & $\Delta$E [MeV] & $E_c$ & $\sum B_{\rm orb}(M11)$ & $\sum B_{\rm spin}(M11)$ & $\sum B_{\rm tot}(M11)$  \\
\hline
exp~\cite{Garotte}              & 1.7 - 3.0 & 2.5 & -    & - & 11.8 \\
sum rules~\cite{Garotte,Enders05}        & -         & 2.1 & 12.1 & - & -\\
QRPA, Gogny~\cite{Garotte,Krt19}  & 1.8 - 2.9  & 2.4 & 9.2 & - & -  \\
GLO~\cite{Garotte,Mump17} & 1.0-6.0 & 3.4 & - & - & 28.2  \\
SMLO~\cite{Garotte,Gor14} & 1.1 - 6.0 & 2.9 & -    & - & 6.3 \\
%E1 simulation~\cite{Garotte}    & 3.0 - 4.7 & 2.5 &      &    &\\
%simulation E1 & 3.0 - 4.7 & 2.5 & - & - & \\
%Oslo (QRPA, Gogny) & 1.8 - 2.9 (?)& 2.4 & 9.0, 9.2 & - & -  \\
WFM~\cite{Balb24} & - & 2.6 & -  & - & 13.4 \\
\hline
SG2  &1.2 - 3.0 & 2.5 & 2.82 & 1.04 &  5.22 \\
 & 1.2 - 4.0 & 2.9 & 6.28 & 1.26 & 8.97 \\
 & 1.2 - 5.0 & 3.8 & 7.47 & 1.59 & 10.19 \\
 \hline
SG2 + J$^2$ & 1.2 - 3.0 & 2.4 & 2.53 & 0.91 & 4.52 \\
 &1.2 - 4.0 & 2.8 & 5.71 & 1.21 & 8.30 \\
 &1.2 - 5.0 & 3.8 & 7.15 & 4.91 & 11.57 \\
\hline
SLy4 & 1.2 - 3.0 & 2.4 & 2.61 & 0.84 & 4.71 \\
&1.2 - 4.0 & 2.9 & 7.77 &  1.22 & 10.31 \\
& 1.2 - 5.0 & 3.1 & 8.87 & 1.40 &  11.65 \\
\hline
SLy5 & 1.2 - 3.0 & 2.5 & 2.62 & 1.44 & 5.72 \\
     &1.2 - 4.0 & 3.0 & 7.36  & 1.92  & 11.03 \\
     & 1.2 - 5.0 & 3.2 & 8.75 & 2.12 & 12.62  \\
\hline
\end{tabular}
\label{tab-4}
\end{table*}

Note that fragmentation of SM strength in $^{254}$No into two groups at $\sim$2.2 MeV
and $\sim$3.8 MeV was also obtained in QRPA calculations with Gogny forces~\cite{Garotte}.
Besides, a fine structure of SM in rare-earth and actinide nuclei was previously found
within QRPA~\cite{Nest_EPJA24,Ne_PRC21,Nest_PAN} and Quasiparticle-Phonon Model (QPM)~\cite{Sol_NPA96}.
where the latter  takes into account the coupling with complex configurations. So
 a fine structure of SM seems to be a common feature of medium and heavy deformed nuclei,
including $^{254}$No. Then the approximation of SM in $^{254}$No by one Lorentzian with centroid
energy 2.5 MeV in experimental analysis~\cite{Garotte} looks questionable.

Further, our calculations do not show any signs of so-called
spin-scissors mode, which following  WFM predictions (see~\cite{Balb_PRC22} and
references therein), should exist in deformed nuclei just below SM. As seen from Fig. 3,
the low-energy spin-flip $M11$ strength is much weaker than the orbital one and the
obtained fine structure of the total strength at 1-5 MeV is mainly explained by
fragmentation of the orbital strength.

For QRPA states in Table~\ref{tab-3}, the summed contribution of two main 2qp
components to the state norm reaches 0.7-0.9. So these states demonstrate a low collectivity
and, as mentioned above, large transition strengths are mainly produced by the
dominant 2qp components. Much more collective states are QRPA excitations
with large $B(E21)$-values, as exhibited in Fig. 3. However, these collective states
are basically quadrupole modes and so do not give large $B(M11)$ values.

In Table~\ref{tab-3}, none of the 2qp configurations corresponds to spin-flip transition.
Instead, they correspond to transitions between the levels arising due to the deformation
splitting, which are known to produce the SM (see the detailed analysis for rare-earth
and actinide nuclei in Ref.~\cite{Ne_PRC21}). We also see a strong constructive interference
of orbital and spin contributions to the total $B_{\rm tot}(M11)$. A similar constructive
interference in SM was previously found in rare-earth and actinide nuclei in the framework
of  QPM~\cite{Sol_NPA96} and self-consistent QRPA~\cite{Nest_EPJA24,Ne_PRC21,Nest_PAN}.

In Table~\ref{tab-4}, our results for SM centroid energies $E_c$ and summed
strengths $\sum B_{\rm orb}(M11)$,  $\sum B_{\rm spin}(M11)$ and $\sum B_{\rm tot}(M11)$
are compared with experimental data~\cite{Garotte} and available previous
estimations. Note that the experiment~\cite{Garotte} cannot separate orbital and
spin $M11$ contributions and so the observed strength should be considered
as the total strength $\sum B_{\rm tot}(M11)$.
Following Table~\ref{tab-4}, the sum rule estimation ~\cite{Garotte,Enders05} roughly reproduces
the experimental data. However, this estimation was done for pure orbital strength
and so, because of the constructive interference of orbital and spin modes in SM energy region,
should be yet significantly increased to obtain $\sum B_{\rm tot}(M11)$.
Further, the QRPA (Gogny) results~\cite{Garotte} look more reasonable  but,
as was mentioned  above, these results were obtained with  a significant ad hoc energy
downshift of the strength~\cite{Krt19}. Next,
%we see from Table~~\ref{tab-4} that
phenomenological models (Generalized Lorentzian (GLO)~\cite{Mump17} and
Simplified Modified Lorentzian  (SMLO)~\cite{Gor14} approximations for GDR)
significantly overestimate or underestimate the summed $M11$ strength.
The WFM~\cite{Balb24} gives  reasonable results but this method is not self-consistent.

Our results in Table~\ref{tab-4} are shown for three energy regions: 1.2-3.0, 1.2-4.0 and 1.2-5.0 MeV
(we have no $1^+$ states below 1 MeV).
These regions include  12, 19, 56 (SG2, SG2+$J^2$) and 9, 17, 35 (SLy4, SLy5)
QRPA states, respectively. The larger density of states for  SG2, SG2+$J^2$ is explained by
the large IS effective mass $m_0/m$=0.79, which should result in denser
single-particle spectra~\cite{Nest_PRC04}.

Following Table~\ref{tab-4}, our calculations for the energy range 1-3 MeV reproduce the
experimental centroid energy $E_c$ but significantly underestimate the experimental
$\sum B_{\rm tot}(M11)$. This is not surprising since in our case (see Fig. 3) the
orbital M11 strength occupies a larger energy interval 1.2-4.0 MeV as compared to the experimental
energy range 1.7-3.0 MeV.  For the
interval 1.2-4.0 MeV, the calculated $\sum B_{\rm tot}(M11)$  is already closer
to the experimental value (especially for SLy5) but still somewhat underestimate it.
At the same time,
we slightly overestimate the centroid energy  $E_c$. For even larger interval 1.2-5.0 MeV,
we reasonably reproduce the experimental $\sum B_{\rm tot}(M11)$ but essentially overestimate
the experimental $E_c$.

In this connection, one should  mention that experimental data~\cite{Garotte} are rather
approximate. Concerning the observed $\sum B(M11)$, the authors write: "It is
important to remark that the former $B(M1,\uparrow)$ value was obtained from those simulations
based on the SLO model for the E1 strength. It is therefore model dependent and should only be
considered as an estimation."  Taking into account the approximate character
of data~\cite{Garotte}, the performance of our calculations looks acceptable.

\subsection{Distributions of the nuclear current}

The low-energy orbital $M11$ strength is called "scissors mode" since, following two-rotor
model~\cite{Iud78}, this flow reminds the IV scissors-like out-of-phase
oscillations of protons and neutron ellipsoids. In this case, the maximal nuclear currents
should take place at the "north" and "south" poles of the deformed nucleus.
However, the nuclear currents for low-energy orbital $M11$ QRPA excitations,
calculated previously for deformed nuclei $^{50}$Cr~\cite{Pai16}
and $^{156}$Gd~\cite{Nest_EPJA24},  show another flow. In these studies, the
maximal currents are obtained rather in the equator sides of the nucleus,
which contradicts the scissors-like treatment. In this connection,  it is
interesting to consider the distributions
of nuclear currents in particular low-energy states of $^{254}$No.

\begin{figure*} %Fig 4
\label{fig4_CTD_2}
\centering
\includegraphics[width=12cm,angle=0, scale=0.85]{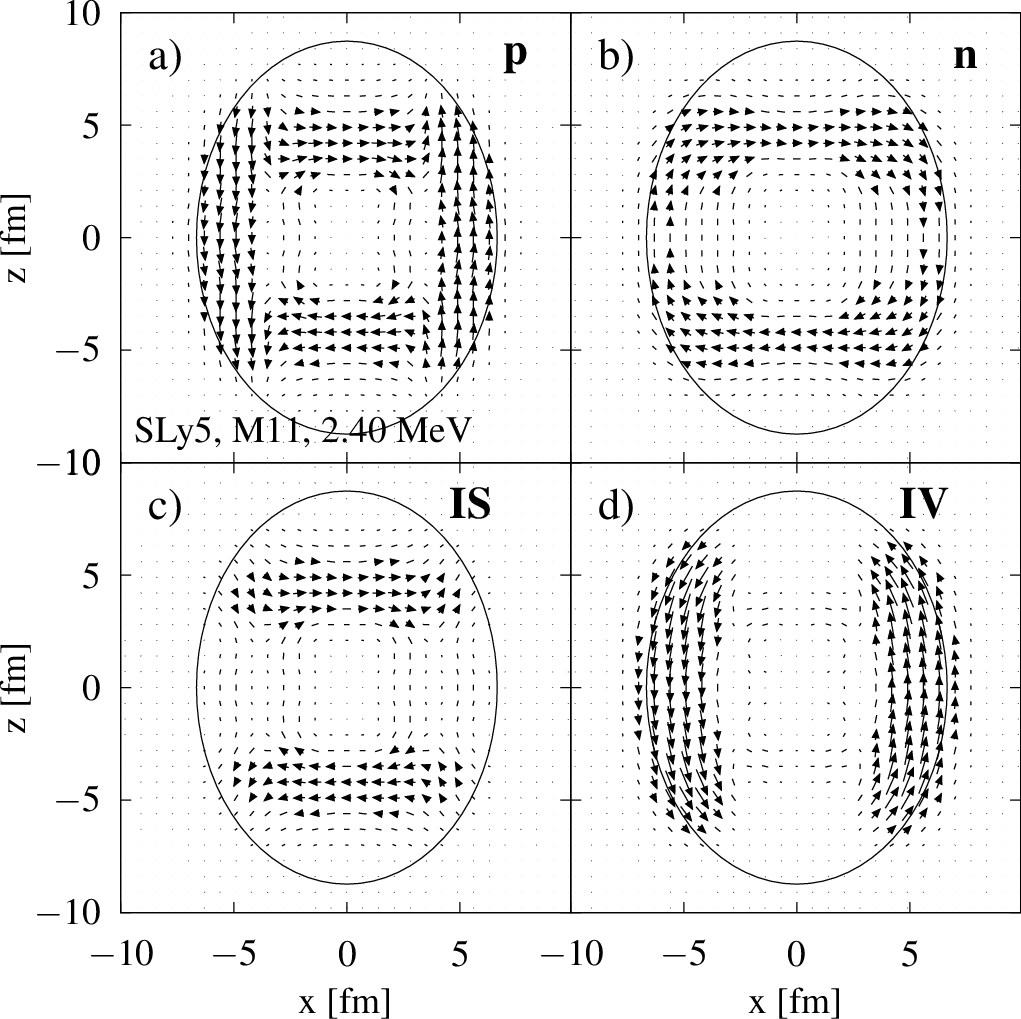}
\caption{Proton (p), neutron (n),  IS and IV CTD for 2.40-MeV state in $^{254}$No, calculated with SLy5.}
\end{figure*}
\begin{figure*} %Fig 5
\label{fig5_CTD_3}
\centering
\includegraphics[width=12cm,angle=0, scale=0.85]{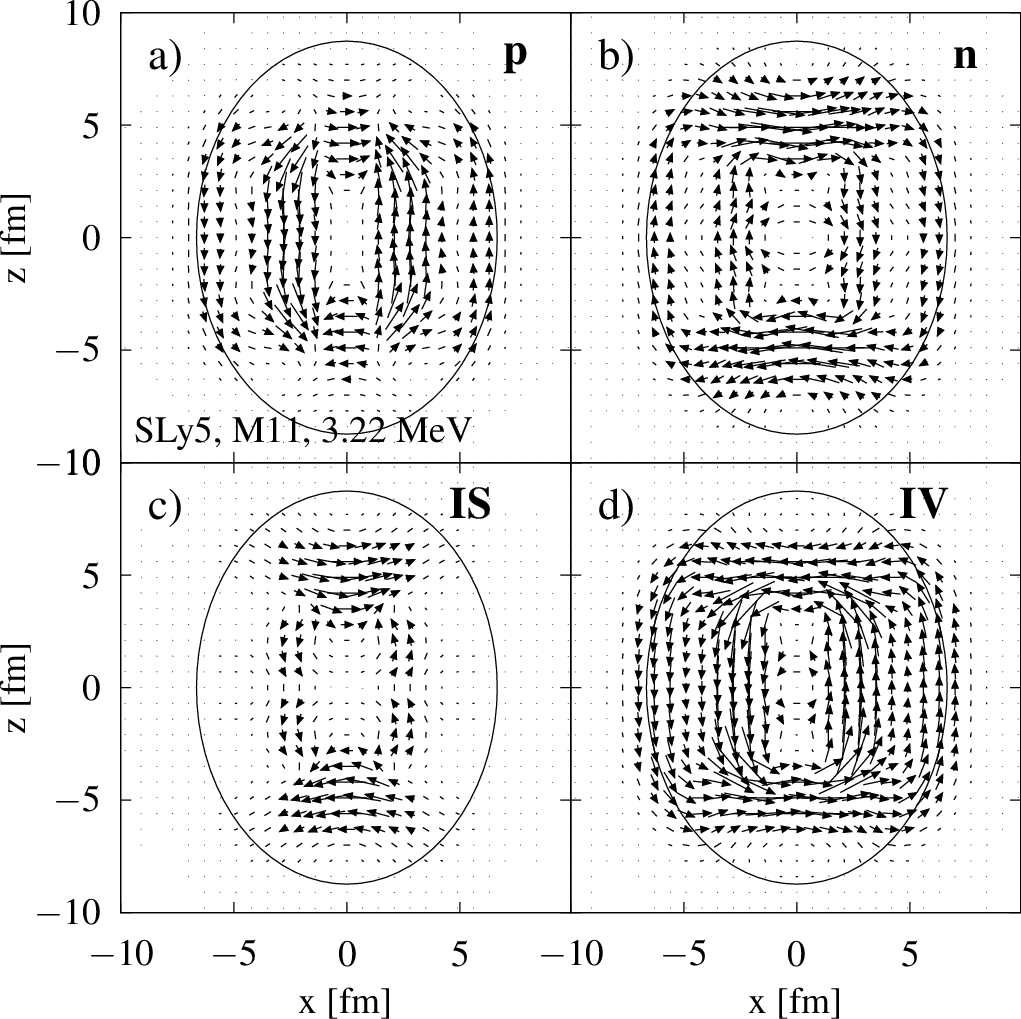}
\caption{The same as in Fig. 4 but for 3.22-MeV state in $^{254}$No, calculated with SLy5.}
\end{figure*}

\begin{figure*} %Fig 6
\label{fig6_IV_CTD}
\centering
\includegraphics[width=12cm,angle=0, scale=0.85]{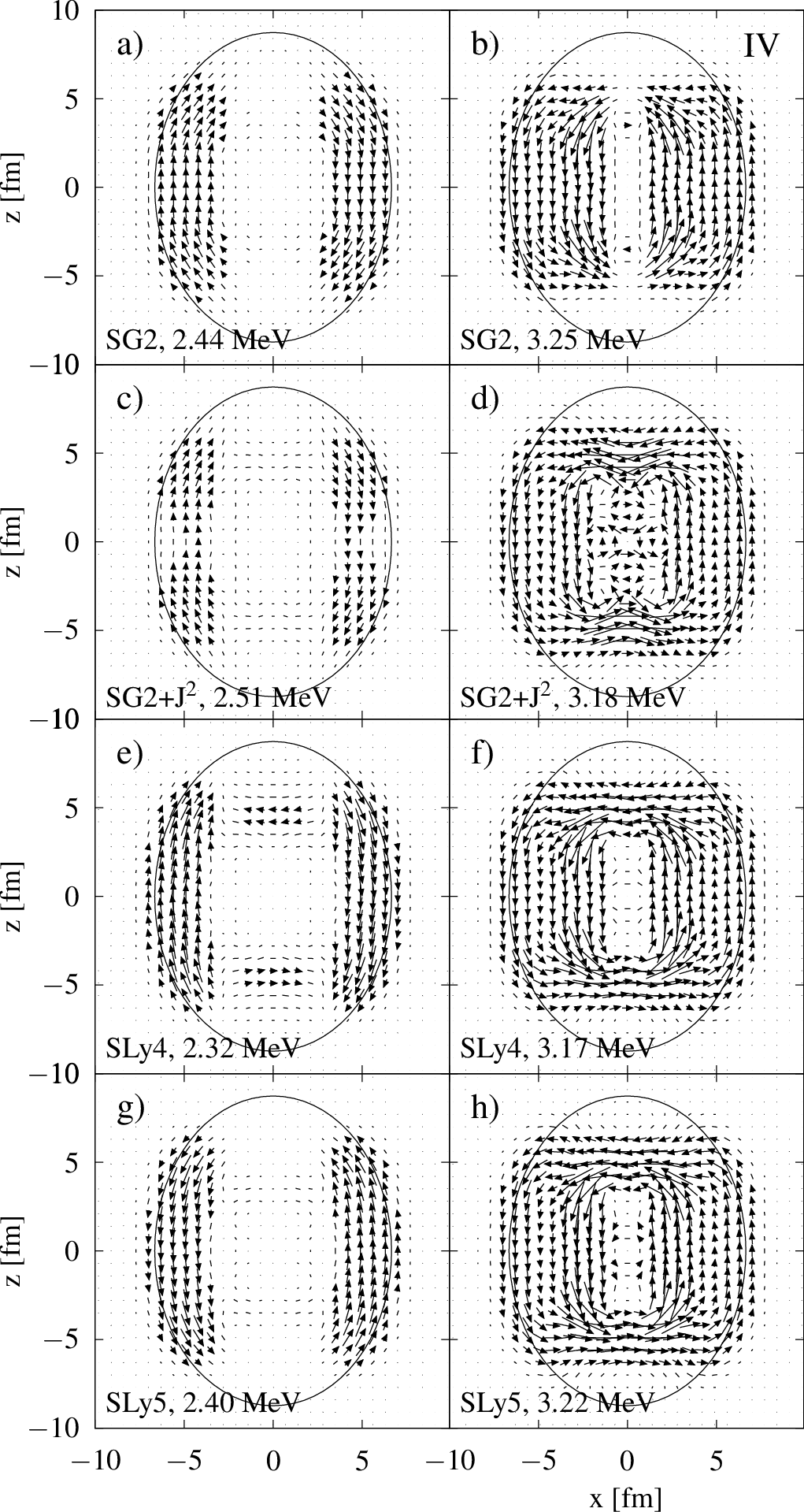}
\caption{IV CTD for some low-energy states in $^{254}$No, calculated with
SG2, SG2+$J^2$, SLy4 and SLy5. }
\end{figure*}

As illustrative examples, we show in Figs. 4-5 the distributions of proton, neutron, IS and IV
nuclear currents for SLy5 2.40-MeV and 3.22-MeV states, listed in Table~\ref{tab-3}.
Like in the previous studies~\cite{Nest_EPJA24,Pai16}, the distributions are given in terms of the
current transition densities (CTD), see Eqs.~(\ref{CTD})-(\ref{j_con}). Only the convective current
is considered.

Figure 4 for 2.40-MeV state shows a small IS current. The IV current dominates but it is mainly
concentrated at the equator left-right sides  of the nucleus,  which contradicts the scissors-like picture.
For 3.22-MeV state in Fig. 5, the IS current is again weaker than  IV one.
%currents are already comparable by the magnitude.
The IS current reminds the quadrupole irrotational flow~\cite{Balb_NPA03}. Besides,
a similar flow arises in the ideal liquid contained in the rotating rigid vessel~\cite{Mikh_PEPAN96}.
The IV current is more complicated than for 2.40-MeV state.
It looks like two clockwise circular flows, where  a  widespread lateral flow
is supplemented by  intense internal flow. Both flows  merge near the pole regions. The IV current again
does not correspond to the scissors-like picture since the current vanishes
 at the nuclear poles.

In Figure 6, IV SLy5 currents are compared with IV currents for SG2, SG2+$J^2$ and SLy4
low-energy states inspected in Table~\ref{tab-3}. Different Skyrme forces give rather similar
flows, which indicates the stability of our results to the choice of the Skyrme parametrization.
The changes with inclusion of $J^2$-term are modest.
A significant difference between CTD for the states at 2.3-2.6 MeV
and 3.1-3.3 MeV  can be explained by different dominant 2qp configurations in these states:
pp[633$\uparrow$, 624$\uparrow$] and pp[521$\uparrow$, 512$\uparrow$], respectively.

\section{Conclusions}
Being inspired by recent experimental data for $M1$ scissors mode (SM) in $^{254}$No~\cite{Garotte},
the low-energy (SM) and higher (spin-flip) $M1(K=1)$  states in $^{254}$No
were analyzed within fully self-consistent Quasiparticle Random-Phase Approximation
(QRPA) model~\cite{Repko_QRPA} based on the Skyrme functional. The Skyrme parametrizations
SG2~\cite{Sg2}, SG2+$J^2$,  SLy4~\cite{SLy5}  and SLy5~\cite{SLy5} were used. The impact
of so-called tensor term  $J^2$ was inspected. This term is absent in SG2 and SLy4 but
introduced by a perturbative way in SG2+$J^2$ and consistently
in SLy5. The deformation-induced coupling between magnetic dipole $M1(K=1)$ and  electric
quadrupole $E2(K=1)$ modes ($M11$ and $E21$ in short notation) was considered.

The Oslo data~\cite{Garotte} describe SM as a bump at 1-4 MeV with the centroid energy
$E_c\sim$2.5 MeV  and summed M1 strength $B_{\rm SM}(M1\uparrow)$=11.8(19) $\mu_N^2$.
Our calculations predict a significant orbital $M11$ strength at 1.2-5.0 MeV. The total (orbital + spin)
$M11$ strength summed in this energy interval (10.2-12.6 $\mu_N^2$) is close
to the experimental value. However, we get much higher centroid energy 3.1-3.8 MeV. Since
the experimental paper~\cite{Garotte} itself considers the obtained data as rather rough estimations,
our agreement with these data looks quite reasonable.

Our calculation reveal that orbital and spin excitation exhibit a significant constructive
interference which leads to an essential increase of the total M1 strength. So, though spin
excitations in SM region are rather weak, their impact through the interference effect is strong.

The $J^2$-term significantly affects spin-flip excitations at 5-10 MeV but is much less
important in SM energy region. This term changes the distribution of M1 strength at 1-5 MeV
(especially between SG2 and SG2+$J^2$ cases) and somewhat increases (by $\sim$ 1 $\mu_N^2$) the
summed M1 strength. As could be expected, the consistent incorporation of $J^2$-term
leads to much smaller changes than its perturbative inclusion. In general, adding the $J^2$-term
gives  a smaller effect than the switch of the Skyrme parametrizations.

The impact of deformation-induced $M11/E21$ coupling in the SM energy region is generally small.
Our calculations do not confirm a hypothesis~\cite{Balb24} on the low-energy spin-scissors resonance.

A fine structure of SM is found. Following our analysis, this structure is mainly caused
by dominant two-quasiparticle components with a large orbital transition probability
$B_{\rm orb}(M11)$.  A distinctive fine structure of SM points out that
approximation~\cite{Garotte} of SM in $^{254}$No by one Lorentzian can be too crude.

The calculated isovector (IV) currents significantly  deviate from the scissors-like flow.
The scissors-like motion should produce maximal currents at the nuclear pole regions but
our calculations predict negligible IV currents there.
A particular behavior of current transition densities seems to be
mainly determined by the dominant two-quasiparticle components of one-phonon states.
 This behaviour is essentially different for states at $\sim$2.5 and  $\sim$3.2 MeV.
 A further analysis of these findings is desirable, e.g. using modern functionals with
 a large IS effective mass.

\section*{Acknowledgement}

The authors thank Prof. P.-G. Reinhard for presentation of code SkyAx and valuable comments.

\appendix
\section{Skyrme functional}
\label{Sec:_Skyrme}

The Skyrme functional has the form \cite{Ves_PRC09,Bender_RMP03}
\begin{eqnarray}
  \mathcal{H}_\mathrm{Sk}
  &=&
  \frac{b_0}{2} \rho^2- \frac{b'_0}{2} \sum\rho_{q}^2
  + \frac{b_3}{3} \rho^{\alpha+2}
  - \frac{b'_3}{3} \rho^{\alpha} \sum \rho^2_q
\nonumber
\\
 &&
 +b_1 (\rho \tau - \textbf{j}^{\;2})
 - b'_1 \sum(\rho_q \tau_q - \textbf{j}^{\;2}_q)
\nonumber
\\
 &&
 - \frac{b_2}{2} \rho\Delta \rho
 + \frac{b'_2}{2} \sum \rho_q \Delta \rho_q
\nonumber
\\
 &&
 - b_4 (\rho \nabla \cdot \textbf{J}\!+\!(\nabla\!\times\!\textbf{j}) \!\cdot\! \textbf{s})
\nonumber
\\
 &&
 - b'_4 \sum (\rho_q \nabla \cdot \textbf{J}_q\!
 +\!(\nabla\!\times\!\textbf{j}_q) \!\cdot\! \textbf{s}_q)
\nonumber
\\
%\end{eqnarray}
%\begin{eqnarray}
  &&
  + \frac{\tilde{b}_0}{2} \textbf{s}^{\;2}
  - \frac{\tilde{b}'_0}{2} \sum \textbf{s}_{q}^{\;2}
+ \frac{\tilde{b}_3}{3} \rho^{\alpha} \textbf{s}^{\;2}
- \frac{\tilde{b}'_3}{3} \rho^{\alpha} \sum \textbf{s}^{\;2}_q
\nonumber
\\
  &&
 -\frac{\tilde{b}_2}{2} \textbf{s} \!\cdot\!
  \Delta \textbf{s} + \frac{\tilde{b}'_2}{2}
  \sum \textbf{s}_q \!\cdot\!\Delta \textbf{s}_q
\nonumber
\\
 &&
  +
  %\gamma_\mathrm{T}
  \tilde{b}_1
   (\textbf{s}\!\cdot\!\textbf{T}\!-\!\textbf{J}^{2})
  +
  %\gamma_\mathrm{T}
  \tilde{b}'_1
   \sum (\textbf{s}_q\!\cdot\!\textbf{T}_q
    \!-\!\textbf{J}_q^{2}) .
\label{eq:skyrme_funct}
\end{eqnarray}
where $b_i$, $b'_i$, $\tilde{b}_i$, $\tilde{b}'_i$ are the functional parameters.
The relation of these parameters with Skyrme standard parameters $t_i$ and $x_i$
can be found elsewhere~\cite{Ves_PRC09,Bender_RMP03,Ne_PRC21}.
Functional (\ref{eq:skyrme_funct}) includes time-even (nucleon
$\rho_q$, kinetic-energy $\tau_q$,
spin-orbit $\textbf{J}_q$) and time-odd (current $\textbf{j}_{ q}$, spin
$\textbf{s}_q$, spin kinetic-energy $\textbf{T}_q$) densities.
The label $q$ denotes protons and neutrons.
Densities without index $q$, like $\rho = \rho_p +
\rho_n$, denote total densities. The contributions with $b_i$ (i=0,1,2,3,4) and
$b'_i$ (i=0,1,2,3) can be treated as isoscalar (IS) and isovector (IV), respectively.
The IV spin-orbit interaction is usually linked to IS
one by $b'_4=b_4$.
The terms with $b_i$ and  $b'_i$ are responsible for ground state
properties and electric excitations of even-even nuclei \cite{Bender_RMP03}.

The terms with $\tilde{b}_i,\tilde{b}'_i$
($i$=0,1,2,3) represent the spin-isospin channel. They are
important for odd nuclei and magnetic modes in even-even nuclei.
The last line of (\ref{eq:skyrme_funct}) includes so-called
 tensor  terms $\propto\tilde{b}_1,\tilde{b}'_1$ which can affect both
 mean-field ground-state properties and magnetic modes. In general, these terms embrace
 contributions from both central-exchange interaction and non-central tensor interaction
 \cite{Colo07,Bender_tensor}. In our calculations, $\textbf{J}^2$-term (depicted as
 $J^2$-term in the main text)  includes only  central-exchange contribution.
In this case, $\tilde{b}_i$ and $\tilde{b}'_i$  are expressed through the standard
Skyrme parameters  $t_1, t_2, x_1, x_2$ as~\cite{Ves_PRC09,Bender_RMP03,Ne_PRC21}
\begin{equation}
 {\tilde b}_1 = \frac{1}{8}(t_1x_1 + t_2x_2), \;
 \tilde{b}'_1 = -\frac{1}{8}(t_1 - t_2) .
 \label{tb1}
 \end{equation}
 For spin-orbit parameter we have
 \begin{equation}
  b_4 = b'_4 = \frac{1}{2}t_4 .
\end{equation}

The concrete values of parameters $ {\tilde b}_1, \tilde{b}'_1$ and $b_4$
for the applied Skyrme parametrizations are shown in Table~\ref{tab-1}
of the main text.


\begin{thebibliography}{99}
\bibitem{Herz08} %1
  R.-D. Herzberg and P. T. Greenlees,
  In-beam and decay spectroscopy of transfermium nuclei,
  Prog. Part. Nucl. Phys. {\bf 61}, 674 (2008).
\bibitem{Hess23} %2
 F.P. Hessberger,
  K isomers in transuranium nuclei,
 arXiv:2309.10468.
\bibitem{Acker25} %3
 D. Ackermann,
 Decay spectroscopy of heavy and superheavy nuclei,
 arXiv:2501.04053v7.
%------------------------------------theory
\bibitem{Dob15} %4
 J. Dobaczewski, A.V. Afanasjev, M. Bender, L.M. Robledo and Yue Shi,
 Properties of nuclei in the nobelium region studied
  within the covariant, Skyrme, and Gogny energy density functionals,
 Nucl. Phys. A {\bf 944}, 388 (2015).
\bibitem{Guo24} %5
 Y. Guo and Y. Shi,
  Stabilization for deformed superheavy nuclei: A shell correction analysis,
  Phys. Rev. C {\bf 110}, 054314 (2024).
\bibitem{Xu24} %6
  F.F. Xu, Y.K. Wang, Y.P. Wang, P. Ring, and P.W. Zhao,
  Emergence of high-order deformation in rotating transfermium nuclei:
   A microscopic understanding,
    Phys. Rev. Lett. {\bf 133}, 022501 (2024).
\bibitem{Dao24} %7
  D.D. Dao and F. Nowacki,
  First complete description of low-lying spectroscopy in $^{254}$No,
  arXiv:2409.08210.
\bibitem{Balb24} %8
 E.B. Balbutsev and I.V. Molodtsova,
 Scissors mode in transuranium elements,
 Eur. Phys. J. A {\bf 60}, 185 (2024).
\bibitem{Nest25} %9
 V.O. Nesterenko, M.A. Mardyban, A. Repko, R.V. Jolos, P.-G. Reinhard and A.A. Dzhioev,
 Low-energy spectra of nobelium isotopes: Skyrme random-phase-approximation analysis,
 Phys. Rev. C {\bf 111}, 064322 (2025).
 %--------------------------experiment
\bibitem{Forge23} %10
  M. Forge et al.,
  New results on the decay spectroscopy of $^{254}$No with GABRIELA@SHELS,
  J. Phys.: Conf. Ser. {\bf 2586}, 012083 (2023).
\bibitem{Wahid25} %11
  S.G. Wahid et al.,
  Isomers and hindrances in $^{254}$No: A touchstone for theories of superheavy nuclei,
   Phys. Rev. C {\bf 111}, 034320 (2025).
\bibitem{Garotte} %12
  F.L. Bello Garrote et al.,
  Experimental observation of the M1 scissors mode in $^{254}$No,
  Phys. Lett. B {\bf 834}, 137479 (2022).
 %---------------------------Scissors Lo Iudice + Lipparini
  \bibitem{Iud78} %13
  N. Lo Iudice and F. Palumbo,
  New isovector collective modes in deformed nuclei,
  Phys. Rev. Lett. {\bf 41}, 1532 (1978).
\bibitem{Boh84} %14  OSR exp
  D. Bohle,  A. Richter, W. Steffen, A.E.L. Dieperink, N. Lo Iudice,
  F. Palumbo, and O. Scholten,
  New magnetic dipole excitation mode studied in the heavy deformed
  nucleus $^{156}$Gd by inelastic electron scattering,
  Phys. Lett. B {\bf 137}, 27 (1984).
 \bibitem{Lip83} %15
 E. Lipparini and S. Stringari,
 Isovector M1 rotational states in deformed nuclei,
 Phys. Lett. B {\bf 130}, 139 (1983).
 \bibitem{LSPR89} %16
 E. Lipparini and S. Stringari,
 Sum rules and giant resonances in nuclei,
 Phys. Rep. {\bf 175}, 103 (1989).
 \bibitem{Hei10} %17
 K. Heyde, P. von Neumann-Cosel, and A. Richter,
 Magnetic dipole excitations in nuclei: Elementary modes of nucleonic motion,
 Rev. Mod. Phys. {\bf 82}, 2365 (2010).
%----------------------------------reviews
 \bibitem{Rich95} %18
   A. Richter,
   Probing the nuclear magnetic dipole response with electrons, photons and hadrons,
   Prog. Part. Nucl. Phys. {\bf 34}, 261 (1995).
 \bibitem{Iud97} %19
  N. Lo Iudice,
  Magnetic dipole excitations in deformed nuclei,
  Phys. Part. Nucl. {\bf 28}, n.6, 1389 (1997).
\bibitem{Enders05} %20
  J. Enders, P. von Neumann-Cosel, C. Rangacharyulu, and  A. Richter,
  Parameter-free description of orbital dipole strength,
  Phys. Rev. C {\bf 71}, 014306 (2005).
%-----------------------------------QRPA-Gogny-MWF
\bibitem{Gogny}%21
  S. Péru and  M. Martini,
  Mean field based calculations with the Gogny force: Some theoretical tools to
  explore the nuclear structure,
  Eur. Phys. J. A {\bf 50}, 88 (2014).
 \bibitem{Mump17} %22
 M.R. Mumpower, T. Kawano, J.L. Ullmann, M. Krticka,  and T. M. Sprouse,
 Estimation of M1 scissors mode strength for deformed nuclei in the medium - to heavy-mass region by
 statistical Hauser-Feshbach model calculations,
 Phys. Rev. C {\bf 96}, 024612 (2017).
 \bibitem{Gor14} %23
 S. Goriely and V. Plujko,
 Simple empirical E1 and M1 strength functions for practical applications,
 Phys. Rev. C {\bf 99}, 014303 (2019).
 %---------------------Skyrme QRPA
 \bibitem{Repko_QRPA} %24
  A. Repko, J. Kvasil, V.O. Nesterenko, P.-G. Reinhard,
  Skyrme RPA for spherical and axially symmetric nuclei,
  arXiv:1510.01248 (nucl-th).
 %-----------------Skyrme forces
 \bibitem{Sg2} %25
  N. Van Giai and H. Sagawa,
  Spin-isospin and pairing properties  of modern Skyrme interactions,
  Phys. Lett. B {\bf 106}, 379 (1981).
 \bibitem{SLy5} %26
   E. Chabanat, P. Bonche, P. Haensel, J. Meyer, R. Schaeffer,
   A Skyrme parametrization from subnuclear to neutron star densities Part II. Nuclei far from stabilities,
   Nucl. Phys. A {\bf 635}, 231 (1998).
 %----------------tensor forces
 \bibitem{Colo07} %27
 G. Colo, H. Sagawa, S. Fracasso and P.F. Bortignon,
 Spin–orbit splitting and the tensor component of the Skyrme interaction,
 Phys. Lett. B {\bf 646}, 227 (2007).
 \bibitem{Ves_PRC09} %28
  P.~Vesely, J.~Kvasil, V.O.~Nesterenko, W.~Kleinig, P.-G.~Reinhard, and  V.Yu.~Ponomarev,
  Skyrme random-phase-approximation description of spin-flip M1 giant resonance,
  Phys. Rev. C {\bf 80}, 031302 (2009).
  \bibitem{Nest_JPG10} %29
 V.O. Nesterenko, J. Kvasil, P. Vesely, W. Kleinig, P.-G. Reinhard, and  V.Yu. Pomomarev,
  Spin-flip M1 giant resonance as a challenge for Skyrme forces,
 J. Phys. G {\bf 37}, 064034, (2010).
  \bibitem{Nest_EPJA24} %30
  V.O. Nesterenko, P.I. Vishnevskiy, P.-G. Reinhard, A. Repko and J. Kvasil,
  Microscopic analysis of dipole electric and magnetic strengths in $^{156}$Gd,
  Eur. Phys. J. A {\bf 60}, 28 (2024).
  %----------------
  \bibitem{Zal_tensor} %31
  M. Zalewski, J. Dobaczewski, W. Satuła, and T.R. Werner,
  Spin-orbit and tensor mean-field effects on spin-orbit splitting including self-consistent
core polarizations,
  Phys. Rev. C {\bf 77}, 024316 (2007).
 \bibitem{Bender_tensor} %32
  M. Bender, K. Bennaceur, T. Duguet, P.-H. Heenen, T. Lesinski, and J. Meyer,
  Tensor part of the Skyrme energy density functional. II. Deformation properties of
magic and semi-magic nuclei,
  Phys. Rev. C {\bf 80}, 064302 (2009).
 \bibitem{Cao_tensor} %33
 Li-Gang Cao, G. Col\`o, H. Sagawa, P.F. Bortignon, and L. Sciacchitano,
  Effects of tensor correlations on low-lying collective states in finite nuclei,
 Phys. Rev. C {\bf 83}, 034324 (2011).
 \bibitem{Pai16} %34
 H. Pai, T. Beck, J. Beller, R. Beyer, M. Bhike, V. Derya, U. Gayer, J. Isaak, Krishichayan, J. Kvasil, B. L¨oher,
 V.O. Nesterenko, N. Pietralla, G. Mart´ınez-Pinedo, L. Mertes, V. Yu. Ponomarev, P.-G. Reinhard, A. Repko,
 P.C. Ries, C. Romig, D. Savran, R. Schwengner, W. Tornow, V.Werner, J.Wilhelmy, A. Zilges, and M. Zweidinger,
 Magnetic dipole excitations of $^{50}$Cr,
 Phys. Rev. C {\bf 93}, 014318 (2016).
 %---------------our QRPA
  \bibitem{Repko_EPJA17_pairing}%35
   A. Repko, J. Kvasil, V.O. Nesterenko, P.-G. Reinhard,
  Pairing and deformation effects in nuclear excitation spectra,
  Eur. Phys. J. A {\bf 53}, 221 (2017).
 \bibitem{Kvasil_EPJA19_SEBRPA} %36
  J. Kvasil, A. Repko, and V.O. Nesterenko,
  Elimination of spurious modes before the solution of quasiparticle
  random-phase-approximation equations,
  Eur. Phys. J.A {\bf 55}, 213 (2019).
\bibitem{Repko_PRC19_SEARPA} %37
   A. Repko, J. Kvasil, and V.O. Nesterenko,
  Elimination of spurious modes within quasiparticle random-phase approximation,
  Phys. Rev. C {\bf 99}, 044307 (2019).
 \bibitem{Bender_RMP03} %38
    M. Bender, P.-H. Heenen, and P.-G. Reinhard,
    Self-consistent mean-field models for nuclear structure,
    Rev. Mod. Phys. {\bf 75}, 121 (2003).
 \bibitem{Nest_PRC04} %39
  V.O. Nesterenko, V.P. Likhachev, P.-G. Reinhard, V.V. Pashkevich, W. Kleinig, and J. Mesa,
  Momentum distribution in heavy deformed nuclei: Role of effective mass,
 Phys. Rev. C {\bf 70}, 057304 (2004).
\bibitem{skyax} %40
    P.-G. Reinhard, B. Schuetrumpf, and J.A. Maruhn,
    The Axial Hartree–Fock + BCS Code SkyAx,
    Comp. Phys. Communic. {\bf 258}, 107603 (2021).
%--------------------experiment
\bibitem{Herz01} %41
 R.-D. Herzberg et al.,
 Spectroscopy of transfermium nuclei: $_{102}^{252}$No,
 Phys. Rev. C {\bf 65}, 014303 (2001).
\bibitem{Rei99} %42
 P. Reiter et al.,
 Ground-State Band and Deformation of the Z=102 Isotope $^{254}$No,
 Phys. Rev. Lett. {\bf 82}, 509 (1999).
 \bibitem{nndc} %43
  Database  http://www.nndc.bnl.gov/nudat3
  %/chartNuc.jsp
 %-----------------books
\bibitem{Har01} %44
   M. N. Harakeh and A. van der Woude,
   {\it Giant Resonances} (Clarendon Press, Oxford, 2001).
   %------------------------------
\bibitem{Sol76} %46
   V.G. Soloviev,
  {\it Theory of Atomic Nuclei} (Pergamon Press, Oxford, 1976).
  %-------------------
 \bibitem{Ne_PRC21} %47
  V.O. Nesterenko, P.I. Vishnevskiy, J. Kvasil, A. Repko, and W. Kleinig,
  Microscopic analysis of the low-energy spin and orbital magnetic dipole excitations in deformed nuclei,
  Phys. Rev. C {\bf 103}, 064313 (2021).
\bibitem{Nest_PAN} %48
  V.O. Nesterenko, P.I.Vishnevskiy, A. Repko, J.Kvasil,
  Low-Energy M1 States in Deformed Nuclei: Spin Scissors or Spin-Flip?
  Phys.Atom. Nuclei {\bf 85}, 858 (2022).
 \bibitem{Sol_NPA96} %49
 V.G. Soloviev, A.V. Sushkov, N.Yu. Shirikova,
 Phys. Rev. C {\bf 53}, 1022 (1996).
 %----------------spin scissors
  \bibitem{Balb_PRC22} %50
   E.B. Balbutsev, I.V. Molodtsova, A.V. Sushkov, N.Yu. Shirikova and P. Schuck,
   Spin-isospin structure of the nuclear scissors mode,
   Phys. Rev. C {\bf 105}, 044323 (2022).
 \bibitem{Krt19} %51
   M. Krti$\check c$ka, S. Goriely, S. Hilaire, S. P$\acute e$ru, and S. Valenta,
 Constraints on the dipole photon strength functions from experimental multistep cascade spectra,
 Phys. Rev. C {\bf 99}, 044308 (2019).
 \bibitem{Balb_NPA03} %45
   E.B. Balbutsev and P. Schuck,
   The nuclear scissors mode in a solvable model,
   Nucl. Phys. A {\bf 720}, 293 (2003);
   E. B. Balbutsev and P. Schuck,
   Erratum to: “The nuclear scissors mode in a solvable model” [Nucl. Phys. A 720 (2003) 293],
   Nucl. Phys. A {\bf 728}, 471 (2003).
 \bibitem{Mikh_PEPAN96} %52
  I.N. Mikhailov, Ch. Brian\c{c}on and P. Quentin,
  Studies of collective currents in nuclei,
  Phys. Elem. Part. Atom. Nucl. {\bf 27}, n.2, 303 (1996).
% P. Ring and P. Schuck,
% {\it Nuclear Many Body Problem}(Springer-Verlag, New York, 1980).

 \end{thebibliography}
\end{document}